# Exploring Use and Perceptions of Generative AI Art Tools by Blind Artists

Gayatri Raman and Erin Brady

Indiana University, Indianapolis  {garaman,brady}@iu.edu

The paper explores the intersection of AI art and blindness, as existing AI research has primarily focused on AI art's reception and impact, on sighted artists and consumers. To address this gap, the researcher interviewed six blind artists from various visual art mediums and levels of blindness about the generative AI image platform Midjourney. The participants shared text prompts and discussed their reactions to the generated images with the sighted researcher. The findings highlight blind artists' interest in AI images as a collaborative tool but express concerns about cultural perceptions and labeling of AI-generated art. They also underscore unique challenges, such as potential misunderstandings and stereotypes about blindness leading to exclusion. The study advocates for greater inclusion of blind individuals in AI art, emphasizing the need to address their specific needs and experiences in developing AI art technologies.

## 1 INTRODUCTION AND RELATED WORK

Generative AI tools are now capable of creating artistic representations using text prompts which describe the desired content of the images [1]. Prior research has examined the reactions of sighted artists to these tools, [1,2], but has not considered the perspectives of visually impaired artists. People with visual impairments may benefit from the text-based nature of AI art prompts, but also cannot get visual information about the images that are being generated to ensure that they meet their expectations.

For this work, we interviewed six blind artists about their knowledge about and use of AI art tools, and asked them to generate AI images and discussed their expectations for the output. In this workshop submission, we reflect on how our preliminary findings and continued research might guide future research on accessible image generation for art creation and consumption

## 2 METHODS AND DEMOGRAPHICS

The participants, described in Table 1, were found through their public art-sharing social media accounts (Instagram, Twitter, and Facebook) and professional websites. A total of six artists were interviewed, ranging in gender, age, art medium, and levels of blindness. The interviews were semi-structured and aimed to capture nuanced insights into their creative processes as blind visual artists, and real-time interactions with the technology. The first half of the interview focused on the individual's journey and artistic process within their lived experience of disability. The participants would then share text prompts that the researcher used to generate images from. The researcher provided a description and interpretation when requested by the participant. The final part of the interview was a discussion on the images and the experience of creating them through the AI tool, and what such images meant for the larger world and processes of art creation and consumption. This work was approved by our institution's Human Subjects Review board.

Two participants F1 and F2 were exclusively digital artists while one, F3, dabbled in digital and non-digital painting. M1, F2, F4, and F5 had received professional training in pottery, graphic design, and visual art respectively. All participants except M1 and F4 were simultaneously engaged in non-artistic occupations while actively commercializing their artwork, frequently through commissioned sales. M1 and F4 were professional artists with art creation as their primary full-time occupation.

All participants had broadly heard of generative AI tools with the most popular one being ChatGPT. They had not previously used Midjourney, the platform used in this study. F1 and F4 had used Be My Eyes, which offered AI-based image descriptions. F2 had also experimented with AI ARTA, an app to generate images in preset filters like anime, "manga realism", etc.

| Gender | Age | Disability | Medium |
|---|---|---|---|
| M1 | 44 | Totally Blind | Ceramic, Pottery, sculpture |
| F1 | 38 | Legally Blind - Retinopathy of Prematurity | Digital Art, Ipad |
| F2 | 42 | Legally blind - Central Areolar Choroidal Dystrophy | Digital Art |
| F3 | 34 | Legally blind - Keratoconus; Epilepsy | Painting, Street Art |
| F4 | 52 | Legally Blind - Optic Nerve Hypoplasia, Right Eye Dominant | Mixed Media, Labyrinths |
| F5 | 72 | Legally Blind - Retinitis Pigmentosa with Usher Syndrome (Tunnel Vision); Hearing Impairment | Painting |

**Table 1**: Participant Demographics

## 3 Findings

In this section, we describe themes that emerged from our interviews that relate to the workshop theme of explainability of AI in the context of the arts.

### 3.1 Comparison with AI art

The participants' reactions to potential comparisons between their work and the images generated by the platform varied based on the medium in which they worked, among other factors.

F1 noted that individuals had on many occasions assumed her art to be AI-generated despite several timelapse videos available on her TikTok showing how she was creating her art "from scratch, digitally". She felt that this assumption was rooted more in her choice of digital art as a medium and less in her blindness. F1 emphasized that her decision to work in digital art, stemmed from accessibility features, such as the backlighting of an iPad preventing shadows, and enhanced zoom capabilities. To some sighted viewers, her work was lesser in effort and value, when in fact her process was "nearly the same as painting in a traditional visual medium like oil or water". F2, who was also a trained and professional graphic designer before her gradual vision loss, echoed this sentiment, of her work as a blind digital artist being devalued by comparisons with AI tools, where she felt that she was already having to prove herself blind "enough". The workshop provides an opportunity to explore how the anxiety and frustration that blind digital and non-digital visual artists feel in comparison with AI tools compare to what sighted artists may experience.

However, nearly all other artists felt unfazed by possible comparisons with AI imagery because their work was unique in style and construction and bore a distinct imprint of their artistic sense, and blindness. M1 felt that as a potter, even the 3D-printed forms of AI images would not be able to recreate the impact of his hand-created ceramic or pottery works. The images generated from his prompt to create a hand-shaped mask with braille lettering did not resemble the mask he had created on his own and also did not contain the required braille lettering as seen in Table 2.

### 3.2 Co-creation and Collaboration with AI tools

Participant M1 was interested in using AI tools to evaluate his existing pottery and ceramic creations indicating a possibility for AI creations to find value in tactile and mixed sensory experiences. Similarly, F2 saw potential in using AI tools to improve or 'clean up' her existing artworks to help with saleability. Her current tool of choice for the same purpose was Photoshop, and she considered the possibility of switching to an AI tool if Photoshop became 'difficult to use' as she progressed in her vision loss.

F4 actively sought ways to inform the process of generating images beyond the prompts themselves, asking "How can I submit pictures of labyrinths so that, you know, the collective consciousness knows what a labyrinth looks like and what guide dog equipment looks like?" M1 felt that "the idea of something is the true creativity" and that the AI tool's work was more "technical than creative". F3 felt that the "open-mindedness" of the images allowed her to think of ways to conceptualize new art pieces and served as a kind of inspiration.

| Prompt from Participant | Images Generated from MidJourney | Limitations |
|---|---|---|
| *"mask in the shape of a human hand and on the back of the hand there will be a human face the fingers will be open on the hand and it will be like slender long fingers like a musician's hands. There will be a ring on each finger, and on the rings there are braille letters. The letters will spell, the word dream. The pinky finger will have D on the ring, the ring finger will have R on the ring, the middle finger will have E, Index finger will have A and the thumb will have M on the ring. Right below the knuckles there will be eyebrows and then the eyes will be partially closed and the nose will be small. The mouth will be small and below the chin, where the wrist is we will put a symbol of a peace sign and the expression on the face will be someone who is just about to wake up from a deep sleep. It will be a left hand."* **M1.** | 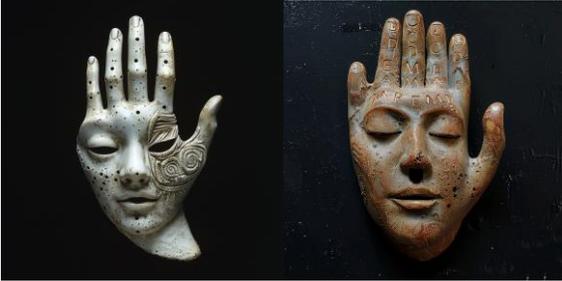 | Inaccurate representation of Braille lettering. |
| *"a mouse reading a braille book inside a lightbulb."* **F4.** | 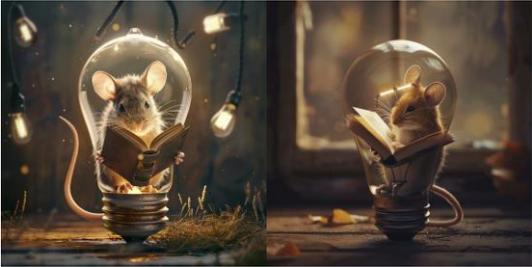 | Only the right image shows the rat possibly feeling the texture of the page, which is how braille is required to be read. The first image is a mouse reading a book regularly, again a misinterpretation of Braille. |
| *"a guide dog and his handler walking through the stars in outer space"* **F4.** | 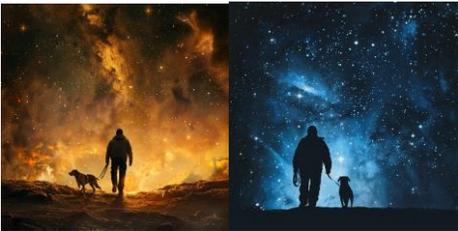 | The images show a regular dog as opposed to a guide dog, as the leash for a guide dog differs from that shown here. |

**Table 2**: Prompts and Images from Image Generation Exercise

*Note.* This table represents a selection of prompts and associated images that the participants suggested which contained Braille or other elements from their life experience of blindness such as a guide dogs or a cane.

F2 and F4 both saw the potential of AI art to become a distinct new medium of art, within which the politics of disability and access to art might emerge as "maybe one day there will be museums dedicated to AI art…photographers didn't replace visual artists". F4 was hopeful about underrepresented communities being able to "put their voice into the imagery pot". She pointed out how the images generated corresponding to the prompt blind man using a guide dog did not accurately depict a guide dog leash, therefore questioning if AI 'understood' the need for guide dogs, and differences from regular dogs and by extension if it understood blindness.

## 3.3 Explainability, 'Visibility' and Blindness

Participants voiced concerns about the potential reinforcement of harmful stereotypes about blindness and the lives of blind individuals through the use of AI image generators like Midjourney. They questioned how these tools were learning and developing 'knowledge' about the world, society, and human behavior, including the experiences of people with disabilities. Artists such as F5, F4, and M1 were not only evaluating the aesthetic quality of the images generated by Midjourney but were also scrutinizing and comparing their creative processes to those of the platform. A critical aspect of their exploration was understanding why the images generated by the platform differed from their own work, even when using prompts that described their original pieces. F4 noted that works at the intersection of language and visual art, such as "ekphrastic poetry" (poetry written about works of art), demonstrated the possibilities within this space of language-assisted art creation.

However, the opaqueness of how the words informed the resulting images was concerning. It restricted artists or creators from engaging with the tool in a way that was artistically meaningful to them. Consequently, participants raised sustained and systemic concerns about the need for access how to the image and prompt repository that the AI trained on or used to develop the resulting images, and how their unique prompts were resulting in the generated images. Based on our findings thus far, there is potential for AI art tools to benefit blind artists, but there were a number of concerns articulated and inaccessible features still left to address. Existing literature on alternative text and captions for images, which aid in explaining visual content on social media, can provide valuable insights. For instance, newer works on AI image captioning, such as "scene descriptions" [3] offer approaches for facilitating a 'conversation' between users and AI. These approaches can help a visually impaired creator gain more creative control over their AI-generated digital art by iteratively modifying the prompt based on information about the contents of each specific image.

The participants' concerns about the image content and the generation processes reflect a deeper need for multifaceted explainability, where artists aim to contextualize the platform's outputs within their own artistic training and life experiences and AI systems can provide the necessary explanations of their outputs for that contextualizing to occur.

## 4 DISCUSSION AND NEXT STEPS

Explainability in the context of what the artists articulated, would be a valuable cognitive tool to unpack the 'creative' process of text-to-image generation in a clear, objective manner. The blind artists in our study sought to understand what happens between the prompt and the generated image, bridging the gap between objective AI processes and the subjective nature of art creation. Therefore, the need for explainability—both objective in linking prompts to images and subjective in revealing perceived artistic choices on the part of the AI tool—remain unmet. Several researchers have debated the extent of potential creativity arising from the work of generative and discriminative neural networks in platforms like Midjourney [2]. Understanding what drives the creative choices of an AI tool involves the algorithm explaining why it made certain decisions, such as the prevalence of certain features or themes in the dataset it was trained on.

By elucidating the inspiration, origins, and interpretations behind its processes, an AI system can make its creative decisions more comprehensible to artists. This level of explainability could help bridge the gap between human creativity and machine-generated art, providing artists with a clearer understanding of the similarities and differences in their creative processes.